\documentclass[aps, prx, reprint, superscriptaddress, longbibliography]{revtex4-1}
\usepackage[colorlinks=true,allcolors=blue]{hyperref}
\usepackage{xcolor}
\usepackage{graphicx}
\usepackage{tikz}
\usepackage{amsmath}
\usepackage{multirow}
\usepackage{rotating}
\usepackage{enumitem}
\usepackage{braket}
\usepackage{changes}

\begin{document} 

\title{Quantum Monte Carlo assessment of embedding for strongly correlated defects: interplay between mean-field starting point and interactions}

\author{Kevin G. Kleiner}
\affiliation{Anthony J. Leggett Institute for Condensed Matter Physics, Department of Physics, Grainger College of Engineering, University of Illinois at Urbana-Champaign, Urbana, Illinois 61801, USA}
\author{Sonali Joshi}
\affiliation{Anthony J. Leggett Institute for Condensed Matter Physics, Department of Physics, Grainger College of Engineering, University of Illinois at Urbana-Champaign, Urbana, Illinois 61801, USA}
\author{Rohan Joshi}
\affiliation{Anthony J. Leggett Institute for Condensed Matter Physics, Department of Physics, Grainger College of Engineering, University of Illinois at Urbana-Champaign, Urbana, Illinois 61801, USA}
\author{Woncheol Lee}
\affiliation{Materials Department, University of California, Santa Barbara, California 93106-5050, USA}
\author{Alexander Hampel}
\affiliation{Center for Computational Quantum Physics, Flatiron Institute, 162 5th Avenue, New York, New York 10010, USA}
\author{Malte R\"osner}
\affiliation{Faculty of Physics, Bielefeld University, 33501 Bielefeld, Germany}
\author{Cyrus E. Dreyer}
\affiliation{Center for Computational Quantum Physics, Flatiron Institute, 162 5th Avenue, New York, New York 10010, USA}
\affiliation{Department of Physics and Astronomy, Stony Brook University, Stony Brook, New York 11794-3800, USA}
\author{Lucas K. Wagner}
\affiliation{Anthony J. Leggett Institute for Condensed Matter Physics, Department of Physics, Grainger College of Engineering, University of Illinois at Urbana-Champaign, Urbana, Illinois 61801, USA}
\date{\today}

\begin{abstract} 
Point defects are of interest for many applications, from quantum sensing to modifying bulk properties of materials. 
Because of their localized orbitals, the electronic states are often strongly correlated, which has led to a proliferation of quantum embedding techniques to treat this correlation.
In these techniques, most of the one-body states are treated with a weakly correlated theory such as density functional theory, and certain one-body states are singled out as an active space to be treated using an effective interaction. 
We assess these techniques for iron and chromium defects in aluminum nitride using quantum Monte Carlo (QMC) calculations on identical Hamiltonians. 
For these systems, we find the dominant errors in the embedding arise from the one-body crystal-field splitting in the d orbitals inherited from density functional theory (DFT), rather than double counting corrections, with the screened interactions also affected by the DFT orbitals. 
Strikingly, the best double counting recipe is opposite in these two cases. 
Because excitation energies can agree while the underlying wave functions do not, diagnosing these errors requires detailed information about the many-body wave functions, which QMC provides. 
\end{abstract}

\maketitle

\section{Introduction}

Point defects, including impurity atoms, lattice imperfections (i.e., vacancies, interstitials, antisites), and complexes involving the two, profoundly affect the properties of materials. 
In electronic devices, point defects can be beneficial, e.g., serving as dopants to tune the conductivity of semiconductors over many orders of magnitude, or detrimental, e.g., trapping charge or causing nonradiative recombination~\cite{shockleyStatisticsRecombinationsHoles1952, hallElectronHoleRecombinationGermanium1952} in optoelectronic devices.
More recently, it has been demonstrated that point defects themselves may serve as quantum devices, including qubits for quantum computing~\cite{weberQuantumComputingDefects2010, kaneSiliconbasedNuclearSpin1998}, single-photon emitters for quantum communication~\cite{aharonovichDiamondbasedSinglephotonEmitters2011, aharonovichSolidstateSinglephotonEmitters2016}, or nanoprobes for quantum metrology~\cite{schirhaglNitrogenVacancyCentersDiamond2014}. 
Recent work has expanded the study of quantum defects to transition metal~\cite{shangFirstprinciplesStudyTransition2022, leeTransitionMetalImpurities2022, otisStronglyCorrelatedStates2025, czelejTransitionMetalRelatedQuantumEmitters2024} and rare-earth~\cite{lvovskyOpticalQuantumMemory2009, kolesovOpticalDetectionSingle2012, awschalomQuantumTechnologiesOptically2018, zhangOpticalSpinCoherence2024} impurities.
Irrespective of the application, it is crucial to have a quantitative understanding of the defect's properties and couplings with the host material.

\textit{Ab initio} computations have served a key role in developing this understanding, with methods based on density functional theory (DFT) serving as the workhorse~\cite{freysoldtFirstprinciplesCalculationsPoint2014}.
However, it has recently become clear that beyond-DFT methods are required to accurately treat several aspects of point defects, especially those needed for quantum applications. 
For example, the treatment of electronic excited states~\cite{mori-sanchezManyelectronSelfinteractionError2006, maExcitedStatesNegatively2010} is necessary since defect qubit initialization and readout procedures commonly use optical excitations and spin-dependent decay pathways~\cite{dohertyNitrogenvacancyColourCentre2013, goldmanStateselectiveIntersystemCrossing2015}; also, excited state properties play a role in non-radiative recombination processes at defects~\cite{alkauskasRoleExcitedStates2016, wickramaratneIronSourceEfficient2016, wickramaratneElectricalOpticalProperties2019}. 
Defects may host strongly-correlated electronic states, i.e., those that have multi-reference character, which cannot be treated with DFT~\cite{vonbarthLocaldensityTheoryMultiplet1979, lischnerFirstPrinciplesCalculationsQuasiparticle2012, muechlerQuantumEmbeddingMethods2022, otisStronglyCorrelatedStates2025}. 
This is especially relevant for impurities with partially-filled orbitals in their one-body band structures~\cite{muechlerQuantumEmbeddingMethods2022, otisStronglyCorrelatedStates2025}.

Quantum embedding methods based on DFT and the constrained random phase approximation (cRPA) have been recently used to address the shortcomings of DFT for defect systems~\cite{bockstedteInitioDescriptionHighly2018, maQuantumSimulationsMaterials2020, maFirstprinciplesStudiesStrongly2020, maQuantumEmbeddingTheory2021, muechlerQuantumEmbeddingMethods2022, otisStronglyCorrelatedStates2025}.
The idea behind these methods is to downfold the full electronic structure of the defect and the host obtained via DFT to a small active space of defect orbitals within which the strongest interactions are assumed to take place. 
The active space properties are described by an effective Hamiltonian with a one-body part taken from DFT and a two-body screened Coulomb interaction from cRPA~\cite{aryasetiawanFrequencydependentLocalInteractions2004, aryasetiawanCalculationsHubbardFirstprinciples2006, miyakeInitioProcedureConstructing2009, sasiogluEffectiveCoulombInteraction2011, pavariniLDA+DMFTApproachStrongly2011}.
These methods commonly assume static two-body interactions and negligible higher-order interactions.
Crucially, the one-body part contains an approximate treatment of the Coulomb interactions in the active space inherited from DFT, so a ``double counting'' (DC) correction is necessary to correctly treat the interactions in the active space~\cite{liechtensteinDensityfunctionalTheoryStrong1995, ryeeEffectDoubleCounting2018}.
DFT+cRPA methods have shown considerable promise, e.g., computing excitation energies for the negatively charged nitrogen-vacancy (NV$^-$) center in diamond and the neutrally-charged silicon-vacancy center in diamond to within a few tenths of an eV from experiment~\cite{maQuantumSimulationsMaterials2020, maFirstprinciplesStudiesStrongly2020, maQuantumEmbeddingTheory2021,muechlerQuantumEmbeddingMethods2022}.

The key challenge for such defect embedding methods is that there are several uncontrolled approximations made when deriving effective Hamiltonians. 
As mentioned above, an approximate DC correction must be chosen, since it is not possible to determine exactly the DFT Coulomb interaction effects in the active space. 
Significant progress in this direction has been made by instead using G$_0$W$_0$ as a starting point for the embedding, in which case the self-energy may be straightforwardly decomposed into contribution inside and outside of the active space~\cite{hauleExactDoubleCounting2015, shengGreensFunctionFormulation2022}.
The use of G$_0$W$_0$ also, to some extent, reduces the dependence on the DFT functional for the initial calculation~\cite{shengGreensFunctionFormulation2022}, though this is expected to be system dependent. 
In addition, depending on the system, the initial electronic configuration used for the DFT or G$_0$W$_0$ calculation may be ambiguous. 
For example, it is common in embedding methods to use a spin-unpolarized mean-field starting point, since exchange interactions are explicitly included in the two-body part of active space Hamiltonian; however, the spin-unpolarized electronic structure for the full system may be very different than the spin-polarized one. 
Finally, the static treatment of interactions in the active space using cRPA is an approximation to be tested~\cite{maQuantumEmbeddingTheory2021,vanloonRandomPhaseApproximation2021}.
Results from embedding methods may depend qualitatively on the choices made for the above approximations.
For example, DFT+cRPA with a DC correction neglecting exchange-correlation effects led to an incorrect low-spin ground state for a neutrally-charged iron substitutional impurity in aluminum nitride (Fe$_{\text{Al}}^0$:AlN)~\cite{muechlerQuantumEmbeddingMethods2022}. 
The G$_0$W$_0$-based scheme of Ref.~\cite{shengGreensFunctionFormulation2022}, on the other hand, gave the correct ground state for Fe$_{\text{Al}}^0$:AlN and lowest excited states within 0.5 eV from experiment~\cite{otisStronglyCorrelatedStates2025}.
The choice of active space orbitals, assumption of static screening in cRPA, and DC correction have each been shown to change the computed excitation energies by at least 0.5 eV for a vanadocene molecule~\cite{changDownfoldingInitioInteracting2024}, which can be considered a model of a strongly correlated defect.

In contrast to embedding methods, \textit{ab initio} quantum Monte Carlo (QMC) methods can provide a full many-body wave function treatment of interactions in defect systems~\cite{foulkesQuantumMonteCarlo2001}.
QMC methods have been commonly used to compute ground state properties of defect systems, including formation energies of Schottky defects in magnesium oxide~\cite{alfeSchottkyDefectFormation2005}, self-interstitials in silicon~\cite{parkerAccuracyQuantumMonte2011}, helium in aluminum~\cite{hoodDiffusionQuantumMonte2012}, oxygen-vacancies in magnesium oxide~\cite{ertekinPointdefectOpticalTransitions2013}, and substitutional nitrogen defects in zinc oxide~\cite{yuFixednodeDiffusionMonte2017}.
Explicit defect excited state calculations with QMC have emerged more recently, with examples including neutrally-charged vacancy centers in diamond~\cite{hoodQuantumMonteCarlo2003}, manganese dopants in phosphors~\cite{saritasExcitationEnergiesLocalized2019}, and the NV$^-$ center in diamond~\cite{chenMulticonfigurationalNatureElectron2023, simulaCalculationEnergiesMultideterminant2023}. 
More recently, some of us introduced an explicit variational principle for excited states in Ref.~\cite{wheelerEnsembleVariationalMonte2024}, offering a means to systematically improve excited state wave functions and obtain ground truth reference results for benchmarking the embedding techniques.

In this work, we compare vertical excitation energies and wave function properties computed with embedding methods against well-controlled, fully \textit{ab initio} QMC calculations for small, periodic AlN supercells containing an Fe$_{\text{Al}}^0$ impurity or a positively-charged chromium impurity (Cr$_{\text{Al}}^+$).
The small supercells are not meant to approximate the dilute defect concentration relevant to compare to experiment, but rather to establish a well-defined comparison between embedding results and ground truth results from QMC.
By comparing results for identical supercells, we are able to isolate errors in the embedding methods and obtain much more detailed information about the many-body wave functions for the full system than is experimentally available.
We find that in these strongly correlated systems, the one-body part of the embedded Hamiltonian is often responsible for the dominant errors, and accurately correcting this via the DC is done with opposite recipes for the two defects. 
Furthermore, the DFT+cRPA procedures lead to overly anisotropic two-body interactions in the active space because of varying orbital localizations from the underlying DFT calculation.
The errors in the embedding procedures are seen to decrease with the larger supercell size, resulting in better quantitative agreement between the embedding and QMC results.

\section{Benchmark systems} \label{section:benchmark_systems}

\subsection{Fe$_{\text{Al}}^0$:AlN} \label{subsection:benchmark_system_FeAlN}

\begin{figure}
\centering
\includegraphics{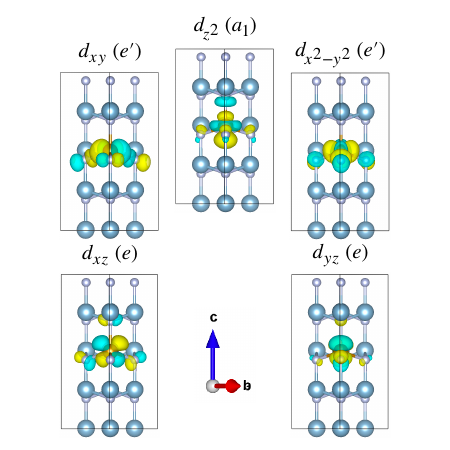}
\caption{Atomic d-like orbitals used for the analysis of Fe$_{\text{Al}}^0$:AlN defect states.
Each orbital is labeled by cubic harmonic character with corresponding irreducible representation (irrep) of the C$_{3v}$ point group, consisting of an e orbital pair, an e' orbital pair, and an a$_1$ orbital. 
Crystallographic vectors $\textbf{a}$, $\textbf{b}$, and $\textbf{c}$ are labeled in the center, with $\textbf{a}$ pointing into the page.
Large blue atoms are Al, small grey atoms are N, and the large orange atom is Fe.}
\label{fig:benchmark_basis_orbitals} 
\end{figure}

\begin{figure}
\centering
\begin{tikzpicture}[x=2cm,y=1cm,thick,>=stealth]

\def\xFA{0}
\def\xTd{1}
\def\xCthreev{2}

\node[align=center] at (\xFA,7) {Free d$^5$ ion};

\draw (\xFA-0.2,1) -- (\xFA+0.2,1);   
\node[below] at (\xFA,1) {$^6$S};

\draw (\xFA-0.2,4.5) -- (\xFA+0.2,4.5);   
\node[below] at (\xFA,4.5) {$^4$G};

\node[align=center] at (\xTd,7) {T$_d$};

\draw (\xTd-0.2,1) -- (\xTd+0.2,1); 
\node[below] at (\xTd,1) {$^6$A$_1$};

\draw (\xTd-0.2,3.2) -- (\xTd+0.2,3.2); 
\node[below] at (\xTd,3.2) {$^4$T$_1$};

\draw (\xTd-0.2,4.5) -- (\xTd+0.2,4.5); 
\node[below] at (\xTd,4.5) {$^4$T$_2$};

\draw (\xTd-0.2,6) -- (\xTd+0.2,6); 
\node[below] at (\xTd,5.8) {$^4$E$/^4$A$_1$};

\node[align=center] at (\xCthreev,7) {C$_{3v}$};

\draw (\xCthreev-0.2,1) -- (\xCthreev+0.2,1); 
\node[right] at (\xCthreev+0.2,1) {$^6$A$_1$};

\draw (\xCthreev-0.2,2.9) -- (\xCthreev+0.2,2.9); 
\node[right] at (\xCthreev+0.2,2.9) {$^4$E};

\draw (\xCthreev-0.2,3.5) -- (\xCthreev+0.2,3.5); 
\node[right] at (\xCthreev+0.2,3.5) {$^4$A$_2$};

\draw (\xCthreev-0.2,4.2) -- (\xCthreev+0.2,4.2); 
\node[right] at (\xCthreev+0.2,4.2) {$^4$E};

\draw (\xCthreev-0.2,4.8) -- (\xCthreev+0.2,4.8); 
\node[right] at (\xCthreev+0.2,4.8) {$^4$A$_1$};

\draw (\xCthreev-0.2,5.7) -- (\xCthreev+0.2,5.7); 
\node[right] at (\xCthreev+0.2,5.7) {$^4$E};

\draw (\xCthreev-0.2,6.3) -- (\xCthreev+0.2,6.3); 
\node[right] at (\xCthreev+0.2,6.3) {$^4$A$_1$};

\draw[dashed,->] (\xFA+0.2,4.5) -- (\xTd-0.2,4.5); 
\draw[dashed,->] (\xFA+0.2,4.5) -- (\xTd-0.2,3.2); 
\draw[dashed,->] (\xFA+0.2,4.5) -- (\xTd-0.2,6);   

\draw[dashed,->] (\xFA+0.2,1) -- (\xTd-0.2,1); 

\draw[dashed,->] (\xTd+0.2,1) -- (\xCthreev-0.2,1);

\draw[dashed,->] (\xTd+0.2,3.2) -- (\xCthreev-0.2,2.9); 
\draw[dashed,->] (\xTd+0.2,3.2) -- (\xCthreev-0.2,3.5); 

\draw[dashed,->] (\xTd+0.2,4.5) -- (\xCthreev-0.2,4.2); 
\draw[dashed,->] (\xTd+0.2,4.5) -- (\xCthreev-0.2,4.8); 

\draw[dashed,->] (\xTd+0.2,6) -- (\xCthreev-0.2,5.7); 
\draw[dashed,->] (\xTd+0.2,6) -- (\xCthreev-0.2,6.3); 

\end{tikzpicture}
\caption{Diagram of energy level splittings of many-body states for a d$^5$ system in ligand field theory without spin-orbit coupling~\cite{suganoMultipletsTransitionMetalIons1970}.
The energy levels for a free d$^5$ ion are split by a T$_d$-symmetric crystal field and a smaller C$_{3v}$-symmetric crystal field.
The CFS is assumed to be smaller than the spherically averaged interaction strength.
This diagram reproduces the same shown in Ref.~\cite{muechlerQuantumEmbeddingMethods2022}.}
\label{fig:d5_ligand_field_theory} 
\end{figure}

The Fe$_{\text{Al}}^0$:AlN defect possesses C$_{3v}$ symmetry with three-fold rotation about the crystallographic $c$-axis and three vertical mirror planes containing this axis~\cite{wickramaratneIronSourceEfficient2016, wickramaratneElectricalOpticalProperties2019, muechlerQuantumEmbeddingMethods2022}.
The four Fe-N bonds form a quasi-tetrahedron with one bond roughly 0.5\% longer than the other three bonds.
Hence, the defect structure has close to T$_d$ (tetrahedral) symmetry.

At the spin-unpolarized DFT level~\cite{wickramaratneIronSourceEfficient2016, wickramaratneElectricalOpticalProperties2019, muechlerQuantumEmbeddingMethods2022}, the system has five Fe d-like orbitals with Kohn-Sham energies in the host band gap, occupied by five particles. 
Based on the C$_{3v}$ crystal field splitting (CFS), the one-body states form a doubly degenerate e orbital pair, doubly degenerate e' orbital pair, and non-degenerate a$_1$ orbital.
Rotations of the five in-gap Kohn-Sham orbitals into atomic Fe d-like orbitals are shown in Fig~\ref{fig:benchmark_basis_orbitals} and labeled using irreducible representations (irreps) of C$_{3v}$.
The a$_1$ and e' Fe d-like orbitals would form the triply degenerate t$_2$ manifold if the site symmetry were T$_d$.
The five d-like orbitals have small lobes around the N sites neighboring the Fe site, indicating finite hybridization with N p orbitals.
We note that this spin-unpolarized DFT electronic structure is very different from the spin-polarized DFT picture~\cite{wickramaratneElectricalOpticalProperties2019}, where the filled majority-spin Fe d bands split from the empty minority spin ones by $\sim 10$ eV, the former ending up below the N p valence band manifold and the latter above the conduction band minimum. 
Thus, the approximations used in the DFT calculation might alter the hybridization of Fe d-like orbitals with host orbitals, so Fe$_{\text{Al}}^0$:AlN is a case where the validity of a spinless DFT starting point for embedding must be confirmed.

From the site symmetry and d occupation of the Fe impurity, we can determine the expected many-body states from ligand field theory and experiments.
The ligand field theory predictions are shown schematically in Fig~\ref{fig:d5_ligand_field_theory}, which reproduces the diagram in Ref.~\cite{muechlerQuantumEmbeddingMethods2022}.
For a free Fe$^{3+}$ ion, the ground state is the spin-5/2 six-fold degenerate $^6$S manifold, while the first excited state is the spin-3/2 36-fold degenerate $^4$G manifold. 
For a relatively small T$_d$ crystal field splitting relative to the Racah interaction parameter $B$, the ground state remains six-fold degenerate, now referred to as $^6$A$_1$. 
With increasing T$_d$ CFS, the $^4$G manifold splits into $^4$T$_1$, $^4$T$_2$, and $^4$E$+^4$A$_1$~\cite{suganoMultipletsTransitionMetalIons1970}, listed in order of increasing many-body energies. 
Lowering the site symmetry from T$_d$ to C$_{3v}$, the $^4$T$_1$ manifold further splits into $^4$A$_2$ and $^4$E, and the $^4$T$_2$ manifold further splits into $^4$A$_1$ and $^4$E.
Low-spin states from the $^2$I manifold of the free ion may also mix with the spin-quadruplet states for larger CFS/$B$; in fact, $^2$T$_2$ becomes the ground state for large CFS/$B$.
Results obtained from emission M\"ossbauer spectroscopy~\cite{masendaLatticeSitesCharge2016} experiments support the small CFS picture with a high-spin ground state.
In addition, photoluminescence experiments obtain a zero-phonon-line (ZPL) of emission at 1.30 eV for this system~\cite{baurDeterminationGaNAlN1994}.

\subsection{Cr$_{\text{Al}}^+$:AlN} \label{subsection:benchmark_system_CrAlN}

\begin{figure}
\centering
\begin{tikzpicture}[x=2cm,y=1cm,thick,>=stealth]

\def\xFA{0}
\def\xTd{1}
\def\xCthreev{2}

\node[align=center] at (\xFA,6.6) {Free d$^2$ ion};

\draw (\xFA-0.2,1.2) -- (\xFA+0.2,1.2);
\node[below] at (\xFA,1.2) {$^3$F};

\draw (\xFA-0.2,3.6) -- (\xFA+0.2,3.6);
\node[below] at (\xFA,3.6) {$^1$D};

\node[align=center] at (\xTd,6.6) {T$_d$};

\draw (\xTd-0.2,1.2) -- (\xTd+0.2,1.2);
\node[below] at (\xTd,1.2) {$^3$A$_2$};

\draw (\xTd-0.2,5.4) -- (\xTd+0.2,5.4);
\node[below=1pt] at (\xTd,5.4) {$^3$T$_2$};

\draw (\xTd-0.2,3.9) -- (\xTd+0.2,3.9);
\node[below] at (\xTd,3.9) {$^1$E};

\node[align=center] at (\xCthreev,6.6) {C$_{3v}$};

\draw (\xCthreev-0.2,1.2) -- (\xCthreev+0.2,1.2);
\node[right] at (\xCthreev+0.2,1.2) {$^3$A$_2$};

\draw (\xCthreev-0.2,3.9) -- (\xCthreev+0.2,3.9);
\node[right] at (\xCthreev+0.2,3.9) {$^1$E};

\draw (\xCthreev-0.2,5.1) -- (\xCthreev+0.2,5.1);
\node[right] at (\xCthreev+0.2,5.1) {$^3$E};

\draw (\xCthreev-0.2,5.7) -- (\xCthreev+0.2,5.7);
\node[right] at (\xCthreev+0.2,5.7) {$^3$A$_1$};

\draw[dashed,->] (\xFA+0.2,1.2) -- (\xTd-0.2,1.2);
\draw[dashed,->] (\xFA+0.2,1.2) -- (\xTd-0.2,5.4);
\draw[dashed,->] (\xFA+0.2,3.6) -- (\xTd-0.2,3.9);

\draw[dashed,->] (\xTd+0.2,1.2) -- (\xCthreev-0.2,1.2);
\draw[dashed,->] (\xTd+0.2,3.9) -- (\xCthreev-0.2,3.9);
\draw[dashed,->] (\xTd+0.2,5.4) -- (\xCthreev-0.2,5.1);
\draw[dashed,->] (\xTd+0.2,5.4) -- (\xCthreev-0.2,5.7);

\end{tikzpicture}
\caption{Diagram of energy level splittings of many-body states for a d$^2$ system in ligand field theory~\cite{suganoMultipletsTransitionMetalIons1970}.
The CFS is assumed to be roughly twice as large as the spherically averaged interaction strength.}
\label{fig:d2_ligand_field_theory} 
\end{figure}

Cr$_{\text{Al}}^+$:AlN has a C$_{3v}$-symmetric geometry and one-body electronic structure qualitatively similar to Fe$_{\text{Al}}^0$:AlN at the spin-polarized DFT level.
The +1 defect charge state (that is, a Cr$^{4+}$ cation) has highest abundance for Fermi energies around mid-gap and below, as shown in prior DFT studies~\cite{czelejTransitionMetalRelatedQuantumEmitters2024, chinnappanFirstprinciplesStudyDefect2025}.
At the spin-polarized DFT level~\cite{wuSynthesisCharacterizationModeling2003, czelejTransitionMetalRelatedQuantumEmitters2024, chinnappanFirstprinciplesStudyDefect2025}, the system has Cr d-like orbitals near the valence band maximum and in the band gap, and the d-like orbitals are occupied by two particles.

For a free Cr$^{4+}$ ion, the ground state is a spin-triplet twenty-one-fold degenerate $^3$F manifold, while the first excited state is a spin-singlet five-fold degenerate $^1$D manifold, as determined from ligand field theory~\cite{suganoMultipletsTransitionMetalIons1970} and shown schematically in Fig~\ref{fig:d2_ligand_field_theory}.
In the T$_d$ crystal field, the ground state degeneracy splits, with the new ground state becoming a spin-triplet three-fold degenerate $^3$A$_2$ manifold.
The first excited states are a spin-triplet nine-fold degenerate $^3$T$_2$ for small CFS or a spin-singlet two-fold degenerate $^1$E for large CFS~\cite{suganoMultipletsTransitionMetalIons1970}.
Existing DFT and quantum chemistry calculations for this system support the large CFS picture with low-spin first excited states~\cite{czelejTransitionMetalRelatedQuantumEmitters2024}.
Lowering the site symmetry from T$_d$ to C$_{3v}$, the $^3$T$_2$ manifold splits into $^3$E and $^3$A$_1$ manifolds.
Photoluminescence experiments measured a zero-phonon-line (ZPL) of emission at 1.20 eV~\cite{baurPhotoluminescenceResidualTransition1995}.

\section{Method} \label{section:method}

\subsection{Geometry relaxation} \label{subsection:geometry_relaxation}

To obtain the ground state equilibrium geometries for Fe$_{\text{Al}}^0$:AlN and Cr$_{\text{Al}}^+$:AlN, we generated $2\times2\times2$ (32-atom) and $3\times3\times2$ (72-atom) supercells of wurtzite AlN using the primitive cell on the Materials Project database~\cite{jainCommentaryMaterialsProject2013}, substituted an Fe atom or a Cr atom for an Al atom, and relaxed the geometries at the spin-unpolarized DFT level.
The DFT calculations used the Perdew-Burke-Ernzerhof (PBE) exchange-correlation functional~\cite{perdewGeneralizedGradientApproximation1996}, a plane wave energy cutoff of 500 eV, projector-augmented wave pseudo-potentials, a $2\times2\times2$ $k$ mesh, and Gaussian thermal smearing with width $\sigma =$ 0.01 eV.
The Fe$_{\text{Al}}^0$:AlN relaxations used \texttt{VASP}~\cite{kresseEfficiencyAbinitioTotal1996}, and the Cr$_{\text{Al}}^+$:AlN relaxations used \texttt{Quantum Espresso}~\cite{giannozziAdvancedCapabilitiesMaterials2017}.
To check consistency between the \texttt{VASP} and \texttt{Quantum Espresso} relaxation procedures, the Fe$_{\text{Al}}^0$:AlN 32-atom cell obtained with \texttt{VASP} was relaxed using \texttt{Quantum Espresso}, and the Fe-N bond lengths were found to change by less than 0.003 \AA{}.

\subsection{\textit{Ab initio} QMC methods} \label{subsection:qmc_methods}

We obtain QMC estimates for the \textit{ab initio} ground state and low-lying excited states at the $\Gamma$ point for the Fe$_{\text{Al}}^0$:AlN and Cr$_{\text{Al}}^+$:AlN supercells.
The QMC calculations employed trial wave functions in multi-Slater-Jastrow form,
\begin{equation}
\Psi(\textbf{R}) = e^{J(\textbf{R})}\sum_kc_k^{}D_k^{\uparrow}(\mathbf{R}_{\uparrow})D_k^{\downarrow}(\mathbf{R}_{\downarrow}),
\label{eqn:msj}
\end{equation}
where $\textbf{R}_{\sigma}$ encompasses the coordinates for all spin-$\sigma$ electrons in the system, $e^J$ is a Jastrow factor, and $D_k^{\uparrow}(\mathbf{R}_{\uparrow})D_k^{\downarrow}(\mathbf{R}_{\downarrow})$ is a Slater determinant written as a product of spin-up and spin-down determinants.

The one-body orbitals were generated from spin-restricted open-shell Kohn-Sham DFT calculations (i.e., for a single set of spatial orbitals for both spin-up and spin-down channels, with spin-dependent occupations) in \texttt{PySCF}~\cite{sunSCFPythonbasedSimulations2018}, using correlation-consistent pseudo-potentials~\cite{annaberdiyevNewGenerationEffective2018}, the PBE0 hybrid exchange-correlation functional~\cite{adamoReliableDensityFunctional1999}, a single $k$ point, and range-separated Gaussian density fitting~\cite{yeFastPeriodicGaussian2021}.
PBE0 combines semi-local correlations with a fraction of Fock exchange in the DFT calculations.
The DFT calculations used a target state of $S_z = +5/2$ for Fe$_{\text{Al}}^0$:AlN or $S_z = +1$ for Cr$_{\text{Al}}^+$:AlN.
For the basis, we retained the highly localized Gaussians describing the atomic cores, while the more diffuse Gaussians were replaced by even-tempered basis functions, providing greater control over the diffuse functions and improved self-consistent field stability for the larger cells.
We checked this basis versus large basis sets for the small systems as described in the supplementary information.

The excited state wave functions were obtained through a Rayleigh-Ritz procedure, in which we expanded the full defect active space in configuration state functions (CSFs); that is, symmetry adapted sums of determinants. 
We then constructed a CSF-Jastrow wave function $\Psi_i(R) = e^{J(R)} C_i(R)$, with $C_i$ the i'th CSF for the given spin and symmetry sector, and $J$ is a fixed Jastrow factor~\cite{wagnerEnergeticsDipoleMoment2007} optimized at the variational Monte Carlo level for the ground state in \texttt{PyQMC}~\cite{wheelerPyQMCAllPythonRealspace2023}.
The Jastrow factor includes one-body electron-nuclear correlations for each nucleus and homogeneous two-body electron-electron correlations.
We then evaluate the Hamiltonian matrix elements $H_{ij} = \langle \Psi_i | \hat{H} | \Psi_j \rangle$ and overlaps $S_{ij} = \langle \Psi_i | \Psi_j \rangle$. 
The resulting matrices were diagonalized using a generalized eigenvalue solver, which provided approximate eigenstates for each spin and symmetry sector. 
Uncertainties were evaluated using a resampling method, which resulted in energy uncertainties of approximately 30 meV. 
The active space for Fe$_{\text{Al}}^0$:AlN included three spin-up and two spin-down electrons in five Fe d-like DFT orbitals, while that for Cr$_{\text{Al}}^+$:AlN included one spin-up and one spin-down electron in five Cr d-like DFT orbitals.
Since the Rayleigh-Ritz procedure is unusual for variational Monte Carlo, we include a workflow in a detailed data repository\cite{kevinkleiner2026data} on the Materials Data Facility.\cite{blaiszikDataEcosystemSupport2019,blaiszikMaterialsDataFacility2016}

To further check the quality of the QMC solutions, we performed diffusion Monte Carlo calculations\cite{foulkesQuantumMonteCarlo2001} on the VMC-generated trial functions. 
For the 1-D irreps, DMC has an upper bound property\cite{foulkesSymmetryConstraintsVariational1999} for those states, although not on the $E_x$ and $E_y$ states. 
Nonetheless, if the nodes of those wave functions are exact, then the diffusion Monte Carlo result will be correct, and we have noted before\cite{pathakExcitedStatesVariational2021} that DMC projection can improve excited state energies. 

\subsection{Quantum embedding methods for defects} \label{subsection:embedding_methods}

The quantum embedding effective Hamiltonians have the form
\begin{align}
\hat{H}_{\text{eff}} = \sum_{i,j \in \mathcal{A}}\sum_{\sigma \in \{\uparrow,\downarrow\}}t_{ij}\left(\hat{c}_{i,\sigma}^{\dagger}\hat{c}_{j,\sigma}^{} + hc\right) \nonumber \\
+ \frac{1}{2}\sum_{i,j,k,l \in \mathcal{A}}\sum_{\sigma,\sigma' \in \{\uparrow,\downarrow\}}U_{ijkl}^{}\hat{c}_{i,\sigma}^{\dagger}\hat{c}_{j,\sigma}^{}\hat{c}_{k,\sigma'}^{\dagger}\hat{c}_{l,\sigma'}^{} - \hat{H}_{\text{DC}},
\label{H_eff}
\end{align}
where $\sigma,\sigma'$ indicate spin-up or spin-down and $i,j,k,l$ index the orbitals in active space $\mathcal{A}$, $t_{ij}$ are one-body hoppings, $U_{ijkl}$ are two-body interactions in chemistry ordering, and $\hat{H}_{\text{DC}}$ is the DC correction.

DFT+cRPA methods require a self-consistent DFT calculation for the Kohn-Sham density, a non-self-consistent DFT calculation for the active space orbitals and one-body hoppings, a cRPA calculation for the screened two-body interactions in the active space, and finally the addition of a DC correction.
The Kohn-Sham densities were computed with spin-unpolarized DFT in \texttt{Quantum Espresso}~\cite{giannozziAdvancedCapabilitiesMaterials2017}, using the PBE functional, plane wave basis with energy cutoff of 5850 eV, correlation-consistent pseudo-potentials consistent with the QMC, $4\times4\times2$ $k$ mesh, and Gaussian thermal smearing with width $\sigma =$ 1.36 $\cdot$ 10$^{-7}$ eV.
The $k$-point mesh was used solely to converge the Kohn-Sham density.
The d-like active space orbitals and hoppings were computed with the same DFT settings, except without self consistency and using a single $k$ point.
To simplify the cRPA calculations, we rotated the five active space orbitals into d-like projected Wannier functions using the \texttt{pw2wannier90} interface to \texttt{Wannier90}~\cite{mostofiUpdatedVersionWannier902014}.
The screened interactions were computed in the Wannier basis using static limit cRPA with 704 bands at the $\Gamma$ point for the 32-atom cells and 784 bands at the $\Gamma$ point for the 72-atom cells, using the \texttt{wan2respack} interface to \texttt{RESPACK}~\cite{nakamuraRESPACKInitioTool2021, kuritaInterfaceToolWannier902023}.
The number of bands was chosen such that the number of unoccupied bands would be the same for the 32- and 72-atom supercell sizes.
Plane wave energy cutoffs of 612 eV were used for all systems.
To check convergence of the plane wave cutoff, we computed excitation energies for the Fe$_{\text{Al}}^0$:AlN 32-atom cell as a function of cutoff setting used, and we found the energies change by roughly 0.05 eV between cutoff settings of 612 eV and 952 eV, as discussed in the Supplemental Material.

We tested the dependence of the DFT+cRPA results on the target state used in the DFT calculation for the full system and on the choice of DC correction in the active space.
The DFT target states involved different occupations of the spin-unpolarized d-like one-body states. 
In our DFT calculations, the states were ordered in energy as e, a$_1$, and e'.
We note the a$_1$ and e' states for Fe$_{\text{Al}}^0$:AlN switch ordering as the supercell size increases~\cite{wickramaratneElectricalOpticalProperties2019, muechlerQuantumEmbeddingMethods2022}. 
For Fe$_{\text{Al}}^0$:AlN, we chose two DFT target states $|22100\rangle$ and $|11111\rangle$, using labels for the total occupations in the five Fe d-like DFT orbitals ordered e, a$_1$, and e'. 
$|22100\rangle$ is the lowest energy state in spin-unpolarized DFT.
For Cr$_{\text{Al}}^+$:AlN, we chose the target state $|11000\rangle$, which represents half filling of the two-fold degenerate e states.
All reference states preserve the C$_{3v}$ symmetry of the defect.

For the DC correction, we considered four cases. 
The first is neglecting the correction altogether. 
At first sight, this seems to be the least rigorous approach; however, it is important to note that in our context, this follows the approach taken by most DFT-based embedding methods. 
The reason is because we are only considering transitions between states in the active space, and the conventional use of the DC correction is to correct the energy spacing between the active space and bulk orbitals. 
In our case, the reason DC makes a difference to the active space spectra is because we are explicitly including \emph{orbital dependence}, which is usually neglected~\cite{liechtensteinDensityfunctionalTheoryStrong1995}.

The second form of the DC we consider is the approximation that the DFT treatment of the Coulomb interaction is described at the Hartree level, but with the cRPA screened interaction~\cite{bockstedteInitioDescriptionHighly2018, maQuantumSimulationsMaterials2020, muechlerQuantumEmbeddingMethods2022}.
Under this assumption, the Hartree DC correction to the one-body parameters $t_{ij}$ has the form 
\begin{equation}
t_{ij}^{\text{Hartree} \; \text{DC}} = \sum_{k,l \in \mathcal{A}}\rho_{kl}U_{ijkl}(\omega = 0),
\label{eqn:hartree_double_counting}
\end{equation}
where $U_{ijkl}(\omega = 0)$ is the interaction tensor in the active space in the static limit and $\rho_{kl}$ is the one-body reduced density matrix in the active space from the DFT calculation.
The Hartree DC correction accounts only for the direct Coulomb interaction and neglects exchange-correlation effects included in the DFT calculation. 
This omission motivates the use of DC corrections that incorporate exchange effects.

The third form of the DC we consider is the Hartree-Fock (HF) approximation to the Coulomb interactions treated in DFT~\cite{bockstedteInitioDescriptionHighly2018}.
In this approach, a Fock exchange term $\frac{1}{2}\rho_{kl}U_{ikjl}$ roughly approximates the exchange contribution to the Kohn-Sham exchange-correlation potential in the active space, thereby partially correcting for self-interaction errors from the use of a Hartree DC correction. 
The HF DC correction is
\begin{equation}
t_{ij}^{\text{HF} \; \text{DC}} = \sum_{k,l \in \mathcal{A}}\rho_{kl}\left(U_{ijkl}(\omega = 0) - \frac{1}{2}U_{ikjl}(\omega = 0)\right).
\label{eqn:hartree_fock_double_counting}
\end{equation}
While the HF DC correction provides a more accurate subtraction than Hartree DC for interactions treated at the DFT level, the fundamental limitation remains that the exact exchange-correlation DC correction is not rigorously defined within DFT+cRPA schemes.
This ambiguity in DC motivates the use of embedding frameworks based on many-body perturbation theory, in which DC corrections can be defined more explicitly.

For the fourth form of the DC, we utilize the G$_0$W$_0$-based embedding method of Ref.~\cite{shengGreensFunctionFormulation2022}. 
This method requires a G$_0$W$_0$ calculation to be performed on the defect supercell after the initial DFT. 
There are two equivalent ways to think about this method. 
It can be considered as an ``exact'' way to include the exchange-correlation contribution in the active space to the DC, i.e., by replacing the exchange-correlation potential by the G$_0$W$_0$ self energy, and then removing the contribution in the active space; the other is simply that the DFT mean-field starting point is replaced by G$_0$W$_0$. 
G$_0$W$_0$-based embedding derives a DC correction of the form

\begin{equation}
t_{ij}^{\text{G}_0\text{W}_0 \; \text{DC}} = V_{ij}^{\text{xc}} + t_{ij}^{\text{Hartree} \; \text{DC}}
- \Sigma_{ij}^{\text{G}_0\text{W}_0}(\omega = 0),
\label{eqn:g0w0_correction}
\end{equation}
where $V_{ij}^{\text{xc}}$ is the Kohn-Sham exchange-correlation matrix element in the active space and $\Sigma_{ij}^{\text{G}_0\text{W}_0}(\omega = 0)$ is the G$_0$W$_0$ self-energy matrix element in the active space in the static limit~\cite{shengGreensFunctionFormulation2022}.
Eq.~\ref{eqn:g0w0_correction} follows from replacing the Kohn-Sham exchange-correlation potential with the G$_0$W$_0$ self-energy within the active space, such that the DC exactly corresponds to the interaction effects included at the G$_0$W$_0$ level. 
Using the DFT eigenvalues and RPA screening, we obtained G$_0$W$_0$ self-energies and DC corrections using the \texttt{pw2bgw} interface to \texttt{BerkeleyGW}~\cite{hybertsenElectronCorrelationSemiconductors1986, deslippeBerkeleyGWMassivelyParallel2012}.
The two-body interactions in the G$_0$W$_0$-based embedding procedure are identical to those derived in DFT+cRPA.

To ensure comparable orbital occupations between the states from DFT+cRPA and QMC, we rotated the final embedded Hamiltonians in the active space from the Wannier basis sets into the reference intrinsic atomic orbital (IAO) basis set, as shown in Fig~\ref{fig:benchmark_basis_orbitals}.
The orbital transformations are nearly one-to-one, as verified in the Supplemental Material.
After rotating, we checked the resulting Hamiltonians are invariant to C$_{3v}$ symmetry operations in the IAO basis.
We then obtained the Hamiltonians' eigenstates using exact diagonalization in~\texttt{PySCF}.

\section{Results} \label{section:results}

\subsection{Reference defect states from QMC} \label{subsection:qmc_data}

We generated approximate eigenstates at the $\Gamma$ point using QMC for the 32- and 72-atom supercells of Fe$_{\text{Al}}^0$:AlN and Cr$_{\text{Al}}^+$:AlN.
If the embedding procedure is correct, the \textit{ab initio} Hamiltonian and embedded Hamiltonians should have low-energy eigenstates with approximately the same excitation energies and qualitatively similar wave functions for the same unit cell.
In this section, we establish the reference QMC results Fe$_{\text{Al}}^0$:AlN and Cr$_{\text{Al}}^+$:AlN.

Strategically, we separate potential errors into two classes.
The first class is solution of the finite unit cell. 
These errors are relevant to the subject of this manuscript and we believe they are small relative to the variance between embedding approaches. 
The second class is the comparison to experiment, which includes finite size effects (experiment is in the dilute limit) and geometrical effects (experiment typically reports the zero phonon line, and we are using density functional theory geometries).

\subsubsection{Finite cell results}

\begin{table}
    \begin{tabular}{lllllll}
\toprule
 & State & $^4$A$_1$ & $^4$A$_2$ & $^4$E$_x$ & $^4$E$_y$  \\
Cell & Method &  &  &  &  &  \\
\hline
\multirow[t]{2}{*}{32} & DMC & 1.93(6) & 1.72(6) & 1.77(5) & 1.83(6) \\
 & VMC & 2.09(4) & 1.75(4) & 1.82(4) & 1.76(4)  \\
\cline{1-7}
\multirow[t]{2}{*}{72} & DMC  & 2.4(2) & 2.5(2) & 1.9(2) & 2.6(2) \\
 & VMC & 2.65(8) & 2.2(1) & 2.30(8) & 2.14(9)  \\
\cline{1-7}
\hline
\end{tabular}
\caption{Excited state energies for the Fe$_{\text{Al}}^0$:AlN defect computed through variational and diffusion Monte Carlo. The $E$ states in the DMC 72-atom cell are not clearly statistically different; we would expect at least one difference to be larger than the error bars.}
\label{table:qmc_fe}
\end{table}

\begin{table*}
\begin{tabular}{lllllllll}
\toprule
 & state & $^1$A$_1$ & $^1$A$_2$ & $^1$E$_x$ & $^1$E$_y$ & $^3$A$_1$ & $^3$E$_x$ & $^3$E$_y$ \\
natom & method &  &  &  &  &  &  &   \\
\hline
\multirow[t]{2}{*}{32} & DMC & 2.75(5) & 3.32(5) & 1.46(5) & 1.47(5) & 2.17(5) &  1.96(5) & 1.78(5) \\
 & VMC & 2.74(4) & 3.58(4) & 1.48(4) & 1.45(4) & 2.56(4) & 2.36(4) & 2.37(4) \\
\cline{1-9}
\multirow[t]{2}{*}{72} & DMC & 2.6(1) & 3.4(1) & 1.4(2) & 1.4(1) & 1.7(1) &  1.9(1) & 1.7(1) \\ 
 & VMC & 2.79(8) & 3.38(9) & 1.54(8) & 1.64(9) & 2.06(9) & 2.06(8) & 1.92(9) \\
\cline{1-9}
\hline
\end{tabular}
\caption{Excited state energies for the Cr$_{\text{Al}}^+$:AlN defect computed through variational and diffusion Monte Carlo.}
\label{table:qmc_cr}
\end{table*}

Tables~\ref{table:qmc_fe} and \ref{table:qmc_cr} show the QMC results for the lowest energy eigenvalue of each symmetry sector. 
An important thing to note is that between VMC and DMC the excitation energies agree to within statistical uncertainties, indicating that the projection does not change the excitation energies (although it does change the total energies significantly). 
We also checked full orbital optimization for a few excitations and saw little change in the excitation energies. 
A more complete analysis will be published elsewhere.
From this information, we thus conclude that the variational excitation energies are reasonably accurate to a level of a few tenths of eV's, which is in line with previous studies on this level of theory.~\cite{changDownfoldingInitioInteracting2024, wagnerDiscoveringCorrelatedFermions2016, pathakExcitedStatesVariational2021, saritasExcitationEnergiesLocalized2019, yuFixednodeDiffusionMonte2017}

 Between the 32-atom and the 72-atom cell, the excitation energies change quite significantly for the Fe$_{\text{Al}}^0$:AlN (around 0.5 eV) while the Cr$_{\text{Al}}^+$:AlN defect has a much smaller size effect. 

\subsubsection{Comparison to ligand field theory}

For the 32- and 72-atom supercells of Fe$_{\text{Al}}^0$:AlN, the QMC calculations obtain a ground state with $^6$A$_1$ symmetry and nearly degenerate lowest three excited states with $^4$A$_2$ and $^4$E symmetries.
The observed ordering of the low-lying eigenstates agrees qualitatively with ligand field theory expectations for a d$^5$ ion with small T$_d$ CFS compared to the Coulomb interactions and a weak T$_d$-to-C$_{3v}$ symmetry lowering~\cite{suganoMultipletsTransitionMetalIons1970}, as illustrated in Fig~\ref{fig:d5_ligand_field_theory}.

The QMC calculations for both Cr$_{\text{Al}}^+$:AlN supercells obtain a ground state with $^3$A$_2$ symmetry, lowest-lying excited states with $^1$E symmetry, and nearly degenerate next lowest-lying excited states with $^3$E and $^3$A$_1$ symmetries.
The ordering of low-lying states from QMC agrees qualitatively with ligand field theory expectations for a d$^2$ ion with large T$_d$ CFS relative to the Coulomb interactions and a weak T$_d$-to-C$_{3v}$ symmetry lowering~\cite{suganoMultipletsTransitionMetalIons1970}, as illustrated in Fig~\ref{fig:d2_ligand_field_theory}.

\subsubsection{Comparison to experiment}

For the Fe$_{\text{Al}}^0$:AlN defect, the $^6$A$_1$ ground state symmetry obtained from QMC coincides with that observed in experiments~\cite{masendaLatticeSitesCharge2016}.
For the excited states, photoluminescence studies report the zero-phonon-line (ZPL) energy of the $^4$E-$^6$A$_1$ transition at 1.30 eV~\cite{baurDeterminationGaNAlN1994}. 
Correcting for lattice relaxation of 0.11 eV~\cite{otisStronglyCorrelatedStates2025}, we obtain a vertical excitation energy estimate of 1.41 eV. 
Our excitation energies are higher than the experimental value, and not getting closer with the larger cell.
The discrepancy may arise from the finite supercell size, geometric effects (as the defect geometry was not relaxed within QMC), effective core potentials, or potentially suboptimal trial wave functions in the QMC calculations. 
It would be interesting in future work to explore whether further finite size scaling improves agreement with experiment; however, for the purposes of this work, we will compare on a finite cell basis between QMC and embedding techniques.

For the Cr$_{\text{Al}}^+$:AlN defect, photoluminescence studies report the ZPL energy of the lowest-lying transition at 1.20 eV~\cite{baurPhotoluminescenceResidualTransition1995}.
Prior DFT calculations found negligible atomic relaxation in the $^1$E first excited states relative to the geometry in the $^3$A$_2$ ground state~\cite{czelejTransitionMetalRelatedQuantumEmitters2024}, enabling a more meaningful comparison (though still approximate) between the QMC-computed vertical excitation energies and the experimental ZPL.
The QMC-computed values for both cell sizes are approximately 0.2 eV higher than experiment, which is consistent with the expected accuracy of state-of-the-art correlated methods.
Because the Cr$_{\text{Al}}^+$:AlN defect appears to have smaller finite size effects, this agreement would be consistent with the hypothesis that QMC energies are accurate and the main limiting factor (compared to experiment) is the finite cell.

\subsection{Comparison of embedding and QMC results} \label{subsection:embedding_qmc_comparison}

In contrast to QMC, DFT+cRPA methods require the user to make several choices for uncontrolled approximations such as the one-body target state in DFT, screening model for cRPA, and DC correction, for which no variational principle exists to guide systematic improvement of the results.
G$_0$W$_0$-based methods seek to remove some of the approximations present in DFT+cRPA with one-body energies based on G$_0$W$_0$ and an exact DC correction~\cite{shengGreensFunctionFormulation2022}.
Here, we investigate which approximations in embedding most affect the computed vertical excitation energies as compared to QMC calculations for the same finite-sized supercell.

\subsubsection{Fe$_{\text{Al}}^0$:AlN} \label{subsubsection:embedding_qmc_comparison_FeAlN}

\begin{figure}
    \includegraphics{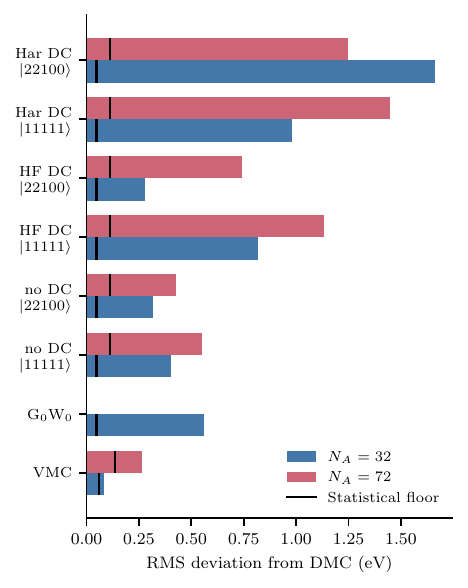}
    \caption{RMS errors of the lowest quartet and sextet energies for the Fe$_{\text{Al}}^0$:AlN defect. 
    $N_A$ is the number of atoms in the cell and the black tick is the expected RMS errors given the statistical uncertainties of the DMC reference.}
    \label{fig:Fe_rms_energy}
\end{figure}

Fig~\ref{fig:Fe_rms_energy} shows the RMS errors of various embedding approaches compared to our DMC results for the Fe$_{\text{Al}}^0$:AlN defect. 
As noted by some of us~\cite{muechlerQuantumEmbeddingMethods2022}, Hartree double counting corrections result in an incorrect ground state, which leads to a large RMS error. 
Hartree-Fock double counting generally improves the agreement, but surprisingly the lowest errors are with no double counting corrections at all. 
Compared to DMC, VMC errors are quite small and nearly within statistical uncertainties.

\begin{figure}
    \includegraphics{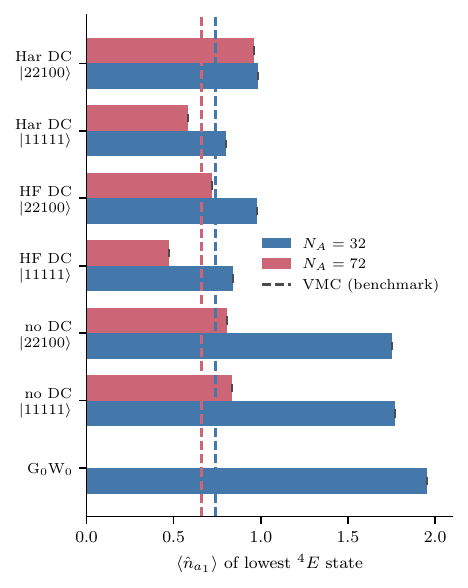}
    \caption{Occupation of the a$_1$ orbital in the $^4$E state for Fe$_{\text{Al}}^0$:AlN defect.
    The VMC result is given by vertical dashed lines for each system size $N_A$. }
    \label{fig:Fe_a1_occupation}
\end{figure}

From the energy errors, one might suppose that the correct recipe for the Fe$_{\text{Al}}^0$:AlN defect is to use no double counting. 
However, even if the energy of the states is correct, the character of the states may still not be correct. 
In Fig~\ref{fig:Fe_a1_occupation}, we show the occupation of the symmetry-distinct a$_1$ orbital in the $^4$E state. 
Even though the "no DC" approach obtains reasonably accurate energies, the states are very different from the QMC reference. 
The large deviation is present in the 32-atom cell, but not in the 72-atom cell.

\begin{figure}
\centering
\includegraphics{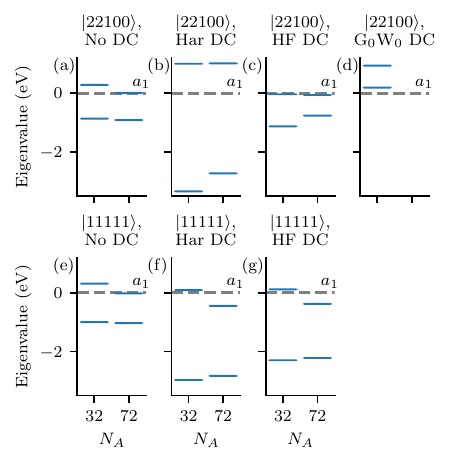}
\caption{e symmetry eigenvalues of the one-body part of the embedded Hamiltonian for the Fe$^0_{\textrm{Al}}$:AlN defect. 
The a$_1$ eigenvalue is set to zero and indicated with a gray dashed line.}
\label{fig:Fe_onebody}
\end{figure}

To understand the behavior of the embedding approaches, in Fig~\ref{fig:Fe_onebody} we show the one-body part of the embedded Hamiltonian for the Fe$_{\text{Al}}^0$:AlN defect.
The Hartree DC corrections decrease the e orbital energies by roughly 1.5 eV more than a$_1$ and e'.
The e orbital energies shift the most because the Fe d$_{xz}$ and d$_{yz}$ orbitals hybridize the least with the neighboring N $p$ orbitals, resulting the higher localization shown in Fig~\ref{fig:benchmark_basis_orbitals}, which then results in a larger Hartree energy and Hartree DC correction.
The enlarged CFS in the Hamiltonian ultimately lead to the low-spin ground state.

The embedded Hamiltonians with HF DC corrections have the correct $^6$A$_1$ ground state.
However, we find the ordering of lowest-lying excited states with the HF DC correction depends sensitively on the DFT target state.
In the DFT $|11111\rangle$ case for the 32-atom cell, for example, the lowest excited state is a spin-doublet $^2$A$_1$, as opposed to the spin-quadruplet $^4$E states in the $|22100\rangle$ case.
In both cases, the HF DC correction reduces the CFS in the Hamiltonian relative to that obtained with Hartree DC correction, as shown in Fig~\ref{fig:Fe_onebody}.
The reduction in CFS resulting from the Fock term is larger for the $|22100\rangle$ case than for the $|11111\rangle$ case.
The effect of the exchange term is largely to cancel the Hartree term  for Fe$_{\text{Al}}^0$:AlN.

\begin{figure}
\centering
\includegraphics{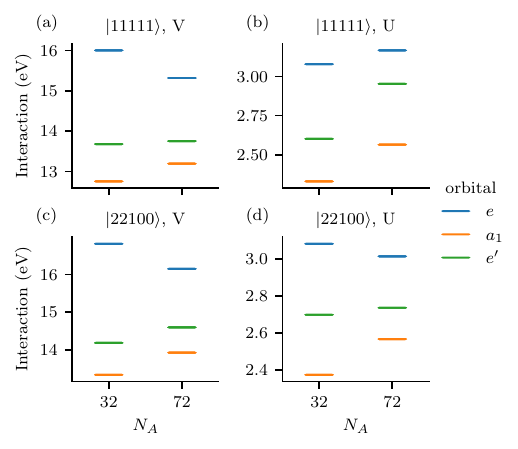}
\caption{Unscreened ($V$) and screened ($U$) interactions for the Fe$_{\text{Al}}^0$:AlN defect. 
Anisotropy in the interactions decreases with larger unit cells.
Since the effect is present in the unscreened interactions, the anisotropy results from orbital anisotropy.}
\label{fig:Fe_interactions}
\end{figure}

Given the strong dependence of the computed spectrum for Fe$_{\text{Al}}^0$:AlN on the CFS in the embedded Hamiltonian, one hypothesis is that G$_0$W$_0$-based embedding procedures with exact G$_0$W$_0$ DC correction might lead to better results than the DFT+cRPA procedures, as was shown in Ref.~\cite{otisStronglyCorrelatedStates2025}.
The Hamiltonian derived with G$_0$W$_0$ has the correct ground state in agreement with the results in Ref.~\cite{otisStronglyCorrelatedStates2025}.
Since the defect has nearly T$_d$ symmetry, we would expect the d orbitals to be split into two degenerate e states at lower energy and three quasi-triply degenerate t$_2$ states at higher energy.
However, as shown in Fig~\ref{fig:Fe_onebody}, the G$_0$W$_0$ one-body energies after DC correction have three nearly degenerate orbitals at lower energy and two at higher energy, which results from a 2 eV T$_d$ CFS from G$_0$W$_0$ combined with an larger correction from the G$_0$W$_0$ DC terms.
The reversal of T$_d$ ordering leads to the correct high-spin ground state and two sets of low-lying $^4$E excited states, thus the high RMS error compared to QMC. 

Regarding the a$_1$ error, ligand field theory provides insight into the origins of the over-occupied a$_1$ orbital.
The ligand field theory models assume a spherically symmetric Coulomb interaction and a one-body CFS that enforces T$_d$ symmetry.
This ligand field theory result coincides qualitatively with the QMC-computed excited states.
See the Supplemental Material for the full ligand field theory analysis for Fe$_{\text{Al}}^0$:AlN.
We infer that cRPA in the 32-atom cell likely overestimates the non-spherical components to the Coulomb interaction in the d orbital active space. 
This can be seen from Fig~\ref{fig:Fe_interactions}, where the intra-orbital screened interaction $U_{a_1}$ is 0.73 eV lower than $U_e$ in the $|22100\rangle$ case and 0.68 eV lower than $U_e$ in the $|11111\rangle$ case; if the interactions were spherically symmetric, $U_{a_1}$, $U_e$, and $U_{e'}$ would be equal.
The bare Coulomb interactions ($V$ terms in Fig~\ref{fig:Fe_interactions}) also have  the same anisotropies, which supports the hypothesis that the PBE calculation obtains large variations in d-like orbital localizations. 
This effect does not appear in the larger 72-atom cell; however, in other defects it could be a source of errors.

\subsubsection{Cr$_{\text{Al}}^+$:AlN} \label{subsubsection:embedding_qmc_comparison_CrAlN}

\begin{figure}
    \includegraphics{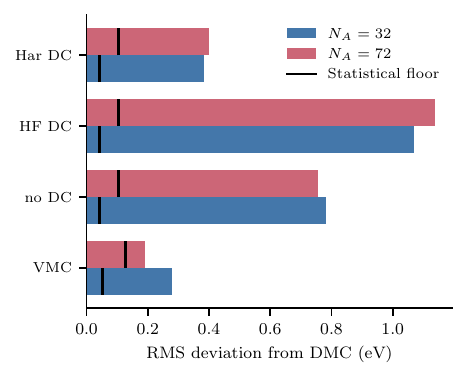}
    \caption{RMS errors of the lowest excitation energies in each symmetry sector for the Cr$_{\text{Al}}^+$:AlN defect. 
    $N_A$ is the number of atoms in the cell and the black tick is the expected RMS errors accounting for statistical uncertainties (``stochastic floor'').}
    \label{fig:Cr_rms_energy}
\end{figure}

\begin{figure}
\centering
\includegraphics{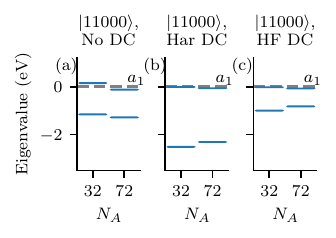}
\caption{e symmetry eigenvalues of the one-body part of the embedded Hamiltonian for the Cr$_{\text{Al}}^+$:AlN defect. 
The gray dashed line represents the a$_1$ state.}
\label{fig:cr_onebody}
\end{figure}

We now compare results between embedding procedures to VMC and DMC for the Cr$_{\text{Al}}^+$:AlN system, which are shown in Fig~\ref{fig:Cr_rms_energy}.
In this case, the obvious initial DFT configuration is $|11000\rangle$; i.e., half filling the two-fold degenerate lowest-energy states, which preserves the C$_{3v}$ symmetry of the defect.
In this case, the Hartree double-counting correction appears to give the most accurate results by some margin, in contrast to the Fe defect. 
The general agreement betwen double-counting approaches is reflected in the one-body energies (Fig~\ref{fig:cr_onebody}), which are qualitatively very similar to one another.
The Hartree double counting correction results in the largest crystal field splitting, which is in the best agreement with the QMC results.

\begin{figure}
    \includegraphics{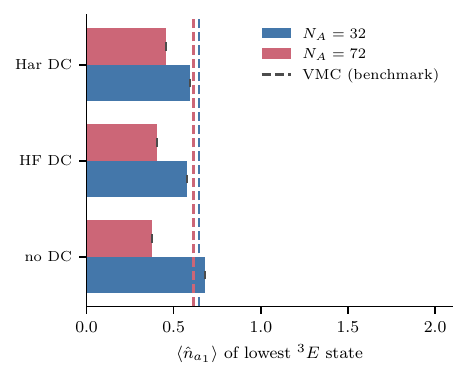}
    \caption{Occupation of the a$_1$ orbital in the $^3$E state for Cr$_{\text{Al}}^+$:AlN  defect.
    The VMC result is given by vertical dashed lines for each system size $N_A$.}
    \label{fig:Cr_a1_occupation}
\end{figure}

As can be noted in Fig~\ref{fig:Cr_a1_occupation}, the errors in the a$_1$ occupation are not as stark in the Cr defect as for the Fe defect. 
This is a result of the fact that the Cr defect has fewer electrons than the Fe defect, which reduces the effects of double counting. 
However, while VMC obtains consistent occupations between the cell sizes, all of the double-counting corrections obtain occupations that are a bit over half the VMC occupation in the 72-atom cell as compared to the 32-atom cell.

\begin{figure}
\centering
\includegraphics{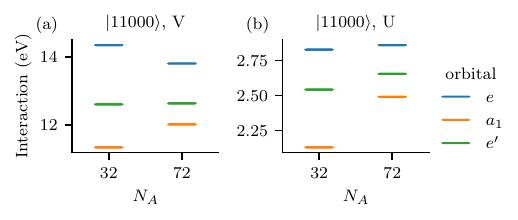}
\caption{Unscreened ($V$) and screened ($U$) for the Cr$_{\text{Al}}^+$:AlN defect. 
The behavior of the anisotropic interactions is similar to the Fe defect.}
\label{fig:cr_interactions}
\end{figure}

We thus can attribute the difference in a$_1$ occupation between embedding techniques and VMC/DMC to the interactions, shown in Fig~\ref{fig:cr_interactions}. 
In this case the key difference is the difference in onsite interactions between the a$_1$ and the e states, which closes with a larger cell. 
This effect appears to be present both in the bare interactions and in the screened interactions, so we attribute this error to \textit{too much} spherical symmetry in the 72-atom case.

\section{Discussion}

We begin the discussion by noting that, especially for the 32-atom cell (but also likely for the 72-atom cell), we have not reached the dilute limit relevant to compare to experiment. 
However, the comparison between QMC and embedding results for each system is well defined and provides insight into the robustness of the embedding procedure. 
Below we will take QMC as the ground truth at each supercell size, and analyze the potential errors in the embedding procedures in detail.

One way of thinking of the DFT+cRPA procedure is as an approximation to the exact embedding construction~\cite{lowdinNoteQuantumMechanicalPerturbation1951, profeExactDownfoldingIts2026}
\begin{equation}
\hat{H}_{\text{eff}}(\omega) = \hat{H}_1(\omega) + \hat{U}(\omega),
\label{eqn:exact_embedding}
\end{equation}
which generally includes explicit frequency dependence in both the one-body Hamiltonian $\hat{H}_1(\omega)$ and interactions $\hat{U}(\omega)$ in the active space.
To aid in the discussion, we classify the errors into the following:
\begin{enumerate}
\item Errors resulting from approximating the one-body Hamiltonian $\hat{H}_1(\omega)$ with the static operator $\hat{H}_{\text{band}} - \hat{H}_{\text{DC}}$, where $\hat{H}_{\text{band}}$ is the band Hamiltonian derived with Kohn-Sham DFT or G$_0$W$_0$ theory.
\item Errors resulting from approximating $\hat{U}(\omega)$ with cRPA. 
\item Errors resulting from neglecting frequency dependence in $\hat{U}(\omega)$. These are expected to be small in our case because of a good separation of levels.
\end{enumerate}

It is important to note that all these error balances can change depending on the one-body states chosen to downfold onto and the underlying Hamiltonian.\cite{changDownfoldingInitioInteracting2024}
In the context of this work, all the embedding active spaces were selected from orbitals computed with the PBE functional in DFT.
Similarly, the computed low-energy eigenstates may be more sensitive to some errors than others in the embedded Hamiltonian.
For the transition metal defects considered, there are well-separated d levels for the defects, which gives some credence to the idea that the cRPA screening model might be sufficiently accurate for the computation of $\hat{U}(\omega)$~\cite{vanloonRandomPhaseApproximation2021}.

\subsection{One-body errors}

The clearest errors originate in the one-body part of the embedded Hamiltonian, and the two defects respond to them in opposite ways.

For Fe$_{\text{Al}}^0$:AlN, no choice of DC corrects the one-body crystal field splitting cleanly. The Hartree DC over-splits the d levels and yields an incorrect low-spin ground state; adding Fock exchange restores the expected high-spin $^6$A$_1$ ground state, but the low-lying spectrum then depends sensitively on the DFT target state.
G$_0$W$_0$-based embedding, which in principle supplies an exact DC, fares no better: its one-body ordering is reversed relative to DFT and ligand field theory for the 32-atom cell~\cite{suganoMultipletsTransitionMetalIons1970}, and its excited states differ qualitatively from QMC.
That even an exact DC fails indicates the error is not primarily one of double counting, but an inherent difficulty in obtaining the correct one-body Hamiltonian from a PBE starting point near the strongly correlated Fe site.
Consistent with this, simply using the PBE band structure with no DC gives the correct ground state and the spectrum closest to QMC.
The part of the DC that actually matters here is its orbital selectivity, which is commonly neglected in embedding of transition-metal systems~\cite{hauleExactDoubleCounting2015, karolakDoubleCountingLDA+DMFTThe2010, kotliarElectronicStructureCalculations2006}.

Cr$_{\text{Al}}^+$:AlN behaves oppositely: the Hartree DC gives the \emph{best} agreement with QMC.
The one-body terms still change substantially with the Hartree DC (Fig~\ref{fig:cr_onebody}), but the low-lying excitations barely probe the crystal field splitting, because the ground and first excited states occupy only the low-energy e orbitals.
The physical reason for the contrast is which orbitals the relevant excitations occupy: Fe's low-lying quartets redistribute charge into the a$_1$ and e' levels and so are exposed to the CFS error, whereas Cr's low-lying excitations are spin flips within the e manifold and are largely insensitive to it.

\subsection{Two-body errors}

The second class of error is a spurious anisotropy in the active-space interactions.
For both defects the intra-orbital interactions $U_{iiii}$, and the bare $V_{iiii}$, differ between the a$_1$, e, and e' orbitals (Figs.~\ref{fig:Fe_interactions} and~\ref{fig:cr_interactions}), whereas a spherically symmetric interaction would make them equal.
Because the effect is already present in the unscreened $V_{iiii}$, it originates in the varying localization of the PBE d-like orbitals rather than in the screening: cRPA screens the interactions but does not correct the underlying orbital shapes.
The anisotropy shrinks in the 72-atom cell, where the PBE orbitals become more spherically symmetric.

As with the one-body error, the consequence depends on the material.
For Fe the low-lying states occupy the a$_1$ orbital, so the anisotropy directly distorts their energies and a$_1$ occupation (Fig~\ref{fig:Fe_a1_occupation}); for Cr the low-lying states avoid a$_1$, so the same anisotropy is largely inconsequential.
This is consistent with the common practice of spherically averaging the interactions or restricting to a single (e or t$_2$) manifold~\cite{pavariniLDA+DMFTApproachStrongly2011}, which removes these errors at the cost of discarding information that, as the Fe case shows, can matter.

Overall, the comparisons suggest that (with no DC), the excited state energies and wave functions from the embedding procedures are more accurate for the 72-atom cells than for the 32-atom cells. 
This improvement is encouraging because the large cell corresponds more to the regime of intended use for defect embedding, i.e., the dilute limit. However, the errors present at small system sizes may inform cases where supercells are large, but with defects that have much more delocalized wave functions.
  
\section{Conclusion} \label{section:conclusion}

We used quantum Monte Carlo (QMC) calculations of excited states to benchmark commonly-used DFT-based embedding techniques for strongly correlated iron and chromium defects in aluminum nitride. 
While these small supercells do not reach the dilute limit relevant to experiment, they enable a like-for-like comparison between full many-body calculations and embedding.
We find that the dominant embedding error is the one-body crystal field splitting inherited from DFT.
The double counting that best corrects the crystal field splitting is opposite for the two defects: none for the iron defect and Hartree double counting for the chromium defect.
Surprisingly, even exact G$_0$W$_0$-based double counting does not correct the crystal field splitting.

The amount of anisotropy in the active-space interaction is a secondary error to the crystal field splitting, and is also inherited from the DFT solution. 
The effects of interaction anisotropy are most evident in the occupation of the a$_1$ orbitals in these defects.
In this case, it is also in the opposite direction in the two defects: in the iron defect the smaller cell is too anisotropic, becoming more isotropic (and closer to QMC) in the larger cell, while in the chromium defect it has the correct anisotropy in the smaller cell, becoming too isotropic in the larger cell.

Finally, our comparisons have shown that excitation energies alone are insufficient to fully determine the correct embedding procedure: embedding methods can match QMC \textit{energies} while producing qualitatively wrong wave functions, as seen in the a$_1$ occupation of the iron $^4$E state. 
Because the wave function errors propagate to derived quantities such as the dipole, electron-phonon couplings, and absorption/emission intensities, many-body techniques such as QMC provide an essential benchmark for defect embedding beyond the energies.

\section{Author Contributions}

\textbf{Kevin G. Kleiner} performed the preliminary set of quantum Monte Carlo calculations and wrote the original draft of the manuscript.
\textbf{Sonali Joshi} aided in analysis of the excited states.
\textbf{Rohan Joshi} and \textbf{Woncheol Lee} performed the embedding calculations.
\textbf{Alexander Hampel}  provided helpful suggestions about the embedding procedures.
\textbf{Malte R\"osner} contributed to revising and writing the manuscript.
\textbf{Cyrus E. Dreyer} and \textbf{Lucas K. Wagner} conceptualized the project, aided in analyzing the results, contributed to the preparation of figures, contributed to writing and revising the manuscript, and supervised the project.
\textbf{Claude Opus (4.7 and 4.8)} were used to generate scripts that generated the symmetry-adapted configuration state functions for QMC. These scripts and their output were carefully checked by a human, as explained in the supplementary information.

\begin{acknowledgments}
CED thanks Coraline Letouz\'{e} for fruitful conversations.
KGK acknowledges support from the National Science Foundation under Award No. DGE-1922758, which supported the preliminary quantum Monte Carlo calculations, and from the NSF Graduate Research Fellowship Program under Award No. DGE-1922758, which supported subsequent calculations and the writing of the original manuscript.
CED acknowledges support from the NSF under Award No. DMR-2237674, which supported the embedding calculations.
KGK, LKW, SJ, and RJ acknowledge support from the U.S. Department of Energy (DOE), Office of Science, Office of Basic Energy Sciences, Computational Materials Sciences Program, under Award No. DE-SC0020177, which supported KGK in writing and revising the manuscript and performing quantum Monte Carlo calculations; RJ in performing embedding calculations; SJ in analyzing data; and LKW in supervising, performing quantum Monte Carlo calculations, writing and revising the manuscript, and preparing the figures.
M.R acknowledges support from the Vidi ENW research programme of the Dutch Research Council (NWO) [Grant DOI: 10.61686/YDRHT18202] with file number VI.Vidi.233.077, which supported editing of the manuscript.
KGK and WL acknowledge support and computing resources from the Flatiron Institute Center for Computational Quantum Physics pre-doctoral program, which supported the preliminary quantum Monte Carlo and embedding calculations shown in the original manuscript.
The Flatiron Institute is a division of the Simons Foundation.
KGK and RJ acknowledge computing resources from the Illinois Campus Cluster Program, which supported additional quantum Monte Carlo and embedding calculations.
An award for computer time was provided by the U.S. DOE's Innovative and Novel Computational Impact on Theory and Experiment (INCITE) Program, which supported additional quantum Monte Carlo calculations.
This research used resources from the Argonne Leadership Computing Facility, a U.S. DOE Office of Science user facility at Argonne National Laboratory, which is supported by the Office of Science of the U.S. DOE under Contract No. DE-AC02-06CH11357.
\end{acknowledgments}

\bibliography{references}

\end{document}